\documentclass[aps,prd,nofootinbib,11pt]{revtex4}
\usepackage{amsfonts}
\usepackage{txfonts}
\usepackage{graphicx}
\usepackage{dcolumn}
\usepackage{bm}
\usepackage{amssymb}
\usepackage{latexsym}
\usepackage[colorlinks, linkcolor=blue, citecolor=blue, urlcolor=red]{hyperref}

\newcommand{\be}{\begin{equation}}
\newcommand{\ee}{\end{equation}}
\newcommand{\bq}{\begin{eqnarray}}
\newcommand{\eq}{\end{eqnarray}}

\begin{document}

\title{Sandage-Loeb test for the new agegraphic and Ricci dark energy models}

\author{Jingfei Zhang}
\affiliation{Department of Physics, College of Sciences,
Northeastern University, Shenyang 110004, China}
\author{Li Zhang}
\affiliation{Department of Physics, College of Sciences,
Northeastern University, Shenyang 110004, China}
\author{Xin Zhang}
\affiliation{Department of Physics, College of Sciences,
Northeastern University, Shenyang 110004, China}

\begin{abstract}

The Sandage-Loeb (SL) test is a unique method to explore dark energy
at the ``redshift desert'' ($2\lesssim z\lesssim 5$), an era not
covered by any other dark energy probes, by directly measuring the
temporal variation of the redshift of quasar (QSO) Lyman-$\alpha$
absorption lines. In this paper, we study the prospects for
constraining the new agegraphic dark energy (NADE) model and the
Ricci dark energy (RDE) model with the SL test. We show that,
assuming only a ten-year survey, the SL test can constrain these two
models with high significance.

\end{abstract}

\keywords{Sandage-Loeb test; new agegraphic dark energy; Ricci dark
energy}

\pacs{98.80.-k, 95.36.+x}

\maketitle

\section{Introduction}\label{sec:introduction}

The astronomical observations of type Ia supernovae (SNIa) indicate
that our universe is undergoing an accelerating expansion~
\cite{Riess98}. This cosmic acceleration has also been confirmed by
other observations, such as the large scale structure
(LSS)~\cite{Tegmark04} and the cosmic microwave background
(CMB)~\cite{Spergel03}. Nowadays it is the most accepted idea that a
mysterious dominant component, dark energy, with large enough
negative pressure, is responsible for this cosmic acceleration.
Among all theoretical models, the preferred one is the so-called
$\Lambda$CDM model, which consists of a mixture of Einstein's
cosmological constant $\Lambda$ and the cold dark matter (CDM). The
$\Lambda$CDM model provides an excellent explanation for the
acceleration of the universe and the existing observational data.
However, the cosmological constant has to face severe theoretical
problems such as the puzzle why the dark energy density today is so
small compared to typical particle scales. Therefore, except the
$\Lambda$CDM model, many dynamical dark energy models have been
proposed, in which the equation of state (EoS) of dark energy is no
longer a constant but slightly evolves with time. For reviews of
dark energy, see, e.g., Ref.~\cite{dark energy}.

In the face of so many candidate models, it is extremely important
to identify which one is the correct model by using the
observational data. The measurement of the expansion rate of the
universe at different redshifts is crucial to discriminate these
competing candidate models. Up to now, a number of cosmological
tools have been used to successfully probe the expansion and the
geometry of the universe. These, typically, include the luminosity
distance of SNIa, the position of acoustic peaks in the CMB angular
power spectrum, and the scale of baryon acoustic oscillations (BAO)
in the power spectrum of matter extracted from galaxy catalogues.
Recently, the application of time evolution of cosmological redshift
as a test of dark energy models has become attractive, since this
method opens a new window of exploring the ``redshift desert''
($2\lesssim z\lesssim 5$). In addition to being a direct probe of
the dynamics of the expansion, this method has the advantage of not
relying on a determination of the absolute luminosity of the
observed sources, but only on the identification of stable spectral
lines, so this method can reduce the uncertainties from systematic
or evolutionary effects.

Sandage \cite{Sandage} was the first to propose the possible
application of this kind of observation as a cosmological tool.
However, owing to the tininess of the expected variation, this
observation was deemed impossible at that time. In 1998, Loeb
\cite{Loeb} revisited this suggestion and argued that the redshift
variation of quasar (QSO) Lyman-$\alpha$ absorption lines could be
detected in the not too distant future, given the advancement in
technology occurred over the last forty years. In fact, the
cosmological redshift variation at 1$\sigma$ would be detected in a
few decades, if a sample of a few hundred QSOs could be observed
with high resolution spectroscopy with a ten meter telescope. This
method is usually referred to as ``Sandage-Loeb'' (SL) test. The
possibility of detecting the temporal variation of redshift with the
Cosmic Dynamics Experiment (CODEX) was first analyzed by Corasaniti,
Huterer and Melchiorri~\cite{Corasaniti:2007bg}. Their
work~\cite{Corasaniti:2007bg} has provided the first quantitative
analysis of the SL test, from which all other analyses have
followed.

In Ref.~\cite{Corasaniti:2007bg}, Corasaniti, Huterer and Melchiorri
employed the SL test to constrain dark energy models such as
$\Lambda$CDM model, Chaplygin gas model, and interacting dark energy
model. Later, Balbi and Quercellini \cite{Balbi:2007fx} extended
this analysis to more dark energy models including constant EoS
model, variable EoS model, interacting dark energy model, DGP model,
Cardassian model, generalized Chaplygin gas model, affine EoS model,
etc. More recently, Zhang, Zhong, Zhu and He \cite{Zhang:2007zg}
further used the SL test to explore the holographic dark energy
model. However, it should be pointed out that there are three
holographic dark energy models: the original holographic dark energy
model \cite{Li:2004rb}, the new agegraphic dark energy model
\cite{nade}, and the Ricci dark energy model \cite{rde}. Actually,
in Ref.~\cite{Zhang:2007zg}, only the original holographic dark
energy model \cite{Li:2004rb}, i.e., the model in which the IR
cutoff is given by the future event horizon, was investigated. Thus,
along this line, as a next step, one should further explore the new
agegraphic and the Ricci dark energy models with the SL test. In
this paper, this will be done. This will provide a complementary to
the work of Ref.~\cite{Zhang:2007zg} and keep the investigation of
holographic dark energy models more complete.

In the subsequent section, we will briefly review the new agegraphic
dark energy model and the Ricci dark energy model. In
Sec.~\ref{sec:test}, we will explore these two models with the SL
test. In the last section, we will give some concluding remarks.

\section{New agegraphic and Ricci dark energy models} \label{sec:agegrapgic}

In this section, we will briefly review the new agegraphic dark
energy model and the Ricci dark energy model. In fact, these two
models both belong to the holographic scenario of dark energy.

It is well known that the holographic principle is an important
result of the recent research for exploring the quantum gravity
\cite{holop}. This principle is enlightened by investigations of the
quantum property of black holes. In a quantum gravity system, the
conventional local quantum field theory will break down because it
contains too many degrees of freedom that would lead to the
formation of a black hole breaking the effectiveness of the quantum
field theory. To reconcile this breakdown with the success of local
quantum field theory in describing observed particle phenomenology,
some authors proposed a relationship between the ultraviolet (UV)
and the infrared (IR) cutoffs due to the limit set by the formation
of a black hole \cite{Cohen99}. The UV-IR relation in turn provides
an upper bound on the zero-point energy density. In other words, if
the quantum zero-point energy density $\rho_{\rm vac}$ is relevant
to a UV cutoff, the total energy of the whole system with size $L$
should not exceed the mass of a black hole of the same size, and
thus we have $L^3\rho_{\rm vac}\leq LM_{\rm Pl}^2$. This means that
the maximum entropy is of the order of $S_{BH}^{3/4}$. When we take
the whole universe into account, the vacuum energy related to this
holographic principle is viewed as dark energy, usually dubbed
holographic dark energy (its density is denoted as $\rho_{\rm de}$
hereafter).

The largest IR cutoff $L$ is chosen by saturating the inequality so
that we get the holographic dark energy density \cite{Li:2004rb}
\begin{equation}
\rho_{\rm de}=3c^2 M_{\rm Pl}^2L^{-2}~,\label{de}
\end{equation} where $c$ is a numerical constant, and
$M_{\rm Pl}\equiv 1/\sqrt{8\pi G}$ is the reduced Planck mass. If we
take $L$ as the size of the current universe, for instance the
Hubble radius $H^{-1}$, then the dark energy density will be close
to the observational result. However, if one takes the Hubble scale
as the IR cutoff, the holographic dark energy seems not to be
capable of leading to an accelerating universe \cite{Hsu04}. The
first viable version of holographic dark energy model was proposed
by Li \cite{Li:2004rb}. In this model, the IR length scale is taken
as the event horizon of the universe. The holographic dark energy
model based on the event horizon as the IR cutoff has been widely
studied \cite{holoext} and found to be consistent with the
observational data \cite{holos09,holofitzx}.

There are also other two versions of the holographic dark energy,
i.e., the new agegraphic dark energy model \cite{nade,nadeext} and
the Ricci dark energy model \cite{rde,Zhang:2009un,ricciext}. For
the new agegraphic dark energy model, the IR scale cutoff is chosen
to be the conformal age of the universe; for the Ricci dark energy
model, the IR cutoff is taken as the average radius of the Ricci
scalar curvature. We shall briefly review these two models in the
following subsections.

\subsection{New agegraphic dark energy model}\label{NADE}

For a spatially flat (the assumption of flatness is motivated by the
inflation scenario) Friedmann-Robertson-Walker (FRW) universe with
matter component $\rho_{\rm m}$ and dark energy component $\rho_{\rm
de}$, the Friedmann equation reads
\begin{equation}\label{Fri}
3M_{\rm Pl}^2H^2=\rho_{\rm m}+\rho_{\rm de}~,
\end{equation}
or equivalently,
\begin{equation}
E(z)\equiv {H(z)\over H_0}=\left(\Omega_{\rm m0}(1+z)^3\over
1-\Omega_{\rm de}\right)^{1/2},\label{Ez}
\end{equation}
where $H\equiv \dot{a}/a$ is the Hubble parameter, $\Omega_{\rm m0}$
is the present fractional matter density, and $\Omega_{\rm de}\equiv
\frac{\rho_{\rm de}}{\rho_{\rm c}} = \frac{\rho_{\rm de}}{3M_{\rm
Pl}^2H^2}$ is the fractional dark energy density.

In the old version of the agegraphic dark energy model
\cite{Cai:2007us}, the IR cutoff is chosen as the age of the
universe $T$ (here it should be pointed out that the light speed has
already been taken as 1, so time and length have the same
dimension). However, there are some inner inconsistencies in this
model; for details see Ref.~\cite{nade}. So, in this paper, we only
discuss the new version of the agegraphic dark energy model. In the
new agegraphic dark energy model, the IR cutoff is chosen to be the
conformal age of the universe,
\begin{equation}
\eta \equiv \int_0^t \frac{dt}{a}=\int_0^a
\frac{da}{a^{2}H}~,\label{etaq}
\end{equation}
so the density of the new agegraphic dark energy is
\begin{equation}
\rho_{\rm de}=3n^{2}M_{\rm Pl}^{2}\eta^{-2}.\label{rhoq}
\end{equation}
To distinguish from the original holographic dark energy model, a
new constant parameter $n$ is used to replace the former parameter
$c$. Taking derivative for Eq. (\ref{rhoq}) with respect to $x=\ln
a$ and making use of Eq. (\ref{etaq}), we get
\begin{equation}
\rho _{\rm de}'= -2 \rho_{\rm de}\frac{\sqrt{\Omega _{\rm de}}}{n
a}. \label{Derivative2}
\end{equation}
This means that the EoS of the new agegraphic dark energy is
\begin{equation}
w_{\rm de}=-1+{2\over 3n}{\sqrt{\Omega_{\rm de}}\over
a}.\label{wade}
\end{equation}
Taking derivative for $\Omega_{\rm de}=n^2/(H^2\eta^2)$, and
considering Eq.~(\ref{etaq}), we obtain
\begin{equation}
\Omega'_{\rm de}=2\Omega_{\rm de}\left(\epsilon-{\sqrt{\Omega_{\rm
de}}\over na}\right),
\end{equation}
where
\begin{equation}
\epsilon={3\over 2}(1+w_{\rm de}\Omega_{\rm de})={3\over 2}-{3\over
2}\Omega_{\rm de}+{\Omega_{\rm de}^{3/2}\over na}.
\end{equation}
Hence, we get the equation of motion for $\Omega _{\rm de}$,
\begin{equation}
\Omega _{\rm de}^{\prime}=\Omega_{\rm de}(1-\Omega_{\rm
de})\left(3-\frac{2}{n}\frac{\sqrt{\Omega_{\rm de}}}{a}\right)~,
\end{equation}
and this equation can be rewritten as
\begin{equation}
\frac{d \Omega_{\rm de}}{dz}=-\Omega_{\rm de}(1-\Omega_{\rm
de})\left(3(1+z)^{-1}-\frac{2}{n}\sqrt{\Omega_{\rm
de}}\right)~\label{keyq}.
\end{equation}
As in Ref.~\cite{nade}, we choose the initial condition,
$\Omega_{\rm de}(z_{\rm ini})=n^2(1+z_{\rm ini})^{-2}/4$, at $z_{\rm
ini}=2000$, then Eq. (\ref{keyq}) can be numerically solved.
Substituting the results of Eq. (\ref{keyq}) into  Eq.~(\ref{Ez}),
the function $E(z)$ can be obtained. Notice that once $n$ is given,
$\Omega_{\rm m0}=1-\Omega_{\rm de}(z=0)$ can be natural obtained by
solving Eq.(\ref{keyq}), so the new agegraphic dark energy model is
a single-parameter model.

\subsection{Ricci dark energy model}\label{RDE}

For a spatially flat FRW universe, the Ricci scalar is
\begin{equation}
{\cal R}=-6\left(\dot{H}+2H^2\right).
\end{equation}
As suggested by Gao et al. \cite{rde}, the energy density of Ricci
dark energy is
\begin{equation}
\rho_{\rm de}={3\alpha\over 8\pi
G}\left(\dot{H}+2H^2\right)=-{\alpha\over 16\pi G}{\cal R},
\end{equation}
where $\alpha$ is a positive numerical constant to be determined by
observations. Comparing to Eq. (\ref{de}), it is seen that if we
identify the IR cutoff $L^{-2}$ with $-{\cal R}/6$, we have
$\alpha=c^2$. As pointed out by Cai et al. \cite{Cai:2008nk}, the
Ricci dark energy can be viewed as originated from taking the causal
connection scale as the IR cutoff in the holographic setting. Now,
the Friedmann equation, in a universe containing Ricci dark energy
and matter, can be written as
\begin{equation}
H^2={8\pi G\over 3}\rho_{\rm m0}e^{-3x}+\alpha\left({1\over
2}{dH^2\over dx}+2H^2\right),
\end{equation}
and this equation can be further rewritten as
\begin{equation}
E^2=\Omega_{\rm m0}e^{-3x} +\alpha\left({1\over 2}{dE^2\over
dx}+2E^2\right),
\end{equation}
where $E\equiv H/H_{0}$. Solving this equation, and using the
initial condition $E_0=E(t_0)=1$, we have
\begin{equation}
E(z)=\left(\frac{2 \Omega_{\rm
m0}}{2-\alpha}(1+z)^{3}+(1-{2\Omega_{\rm m0}\over 2 -\alpha})
(1+z)^{(4-{2\over\alpha})}\right)^{1/2}.\label{Ea}
\end{equation}
There are two model parameters, $\Omega_{\rm m0}$ and $\alpha$, in
the Ricci dark energy model.

\section{The Sandage-Loeb test}\label{sec:test}

In this section, we will first review the Sandage-Loeb test, and
then explore the new agegraphic dark energy model and the Ricci dark
energy model with the SL test.

First, let us consider an isotropic source emitting at rest. The
well-known redshift relation of the radiation emitted by the source
at $t_s$ and observed at $t_o$ is
\begin{equation}
z_s(t_o)=\frac{a(t_o)}{a(t_s)}-1.~\label{zz}
\end{equation}
Furthermore, consider lights emitted after a period $\Delta{t}_s$ at
$t_s+\Delta{t}_s$ and detected later at $t_o+\Delta{t}_o$.
Obviously, the observed redshift of the source at $t_o+\Delta{t}_o$
is
\begin{equation}
z_s(t_o+\Delta t_o)=\frac{a(t_o+\Delta t_o)}{a(t_s+\Delta t_s)}-1.
\end{equation}
Therefore, the variation of the source redshift between times $t_o$
and $t_o+\Delta t_o$ would be measured as follows:
\begin{equation}
\Delta{z}_s\equiv\frac{a(t_o+\Delta t_o)}{a(t_s+\Delta t_s)}-
\frac{a(t_o)}{a(t_s)}.\label{dz}
\end{equation}
We can expand the ratio $a(t_o+\Delta t_o)/a(t_s+\Delta t_ s)$ to
linear order, under the approximation $\Delta t/t\ll 1$.
Furthermore, using the the relation $\Delta
t_o=[a(t_o)/a(t_s)]\Delta t_s$, we obtain
\begin{equation}
\Delta z_s \approx \left[\frac{{\dot a}(t_o)-{\dot a}(t_
s)}{a(t_s)}\right]\Delta t_o.
\end{equation}
It shows that the redshift variation $\Delta z_s$ is directly
related to a change in the expansion rate during the evolution of
the universe, and it is thus a direct probe of the dynamics of the
cosmic expansion. This redshift variation can be related to a
spectroscopic velocity shift, $\Delta v\equiv \Delta z_s
/(1+z_{s})$. Using the Hubble parameter $H(z)=\dot{a}(z)/a(z)$, we
obtain
\begin{equation}
\Delta v=H_0\Delta
t_o\left[1-\frac{E(z_s)}{1+z_s}\right],\label{deltav}
\end{equation}
where $H_0$ is the Hubble constant and $E(z)=H(z)/H_0$. The function
$E(z)$ contains all the details of the cosmological model under
investigation. It is clear that the expansion history $E(z)$ is
related to the spectroscopic velocity shift via Eq.~(\ref{deltav}).

Though the amplitude of the velocity shift is very small, the
absorption lines in the quasar Lyman-$\alpha$ provide us with a
powerful tool to detect such a small signal. Monte Carlo simulations
of Lyman-$\alpha$ absorption lines have been performed to estimate
the uncertainty on $\Delta v$ as measured by the CODEX spectrograph
\cite{Pasquini05}. The statistical error can be estimated as
\begin{equation}
\sigma_{\Delta
v}=1.4\left(\frac{2350}{S/N}\right)\sqrt{\frac{30}{N_{QSO}}}
\left(\frac{5}{1+z_{QSO}}\right)^{1.8}~{{\rm cm}\over {\rm
s}},\label{error}
\end{equation}
where $S/N$ denotes the spectral signal-to-noise defined per 0.0125
{\AA} pixel, $N_{QSO}$ is the number of Lyman-$\alpha$ quasars, and
$z_{QSO}$ is the quasar's redshift. In order to detect the cosmic
signal, a large $S/N$ is necessary, but this implies that a positive
detection is not feasible with current telescopes. Fortunately, the
CODEX under design will be installed on the ESO Extremely Large
Telescope. The necessary signal-to-noise can be reached by such an
about 50 meter giant with just few hours integration. The velocity
shift measurements open a cosmological window with particular focus
on dark energy models. From the velocity shift measurements, one can
forecast constraints on parameters of cosmological models. In this
paper, following
Refs.~\cite{Corasaniti:2007bg,Balbi:2007fx,Zhang:2007zg}, we
consider experimental configuration and uncertainties similar to
those expected from CODEX. We assume that the survey would observe a
total of 240 QSOs uniformly distributed in six equally spaced
redshift bins in the range $2\lesssim z\lesssim 5$, with a
signal-to-noise $S/N=3000$, and the expected uncertainty as given by
Eq.~(\ref{error}). Also, in this paper, we consider a ten-year
survey, namely, $\Delta t_o=10$ years. In what follows, we shall
explore the new agegraphic dark energy model and the Ricci dark
energy model with the SL test.


\begin{figure}[htbp]
\centering $\begin{array}{c@{\hspace{0.2in}}c}
\multicolumn{1}{l}{\mbox{}} &
\multicolumn{1}{l}{\mbox{}} \\
\includegraphics[scale=0.8]{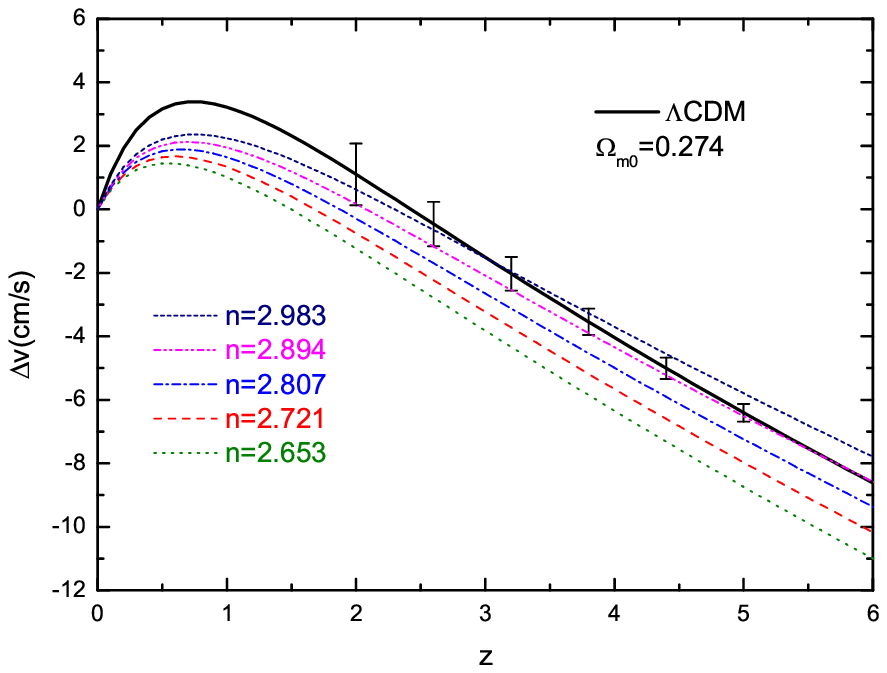} &\includegraphics[scale=0.8]{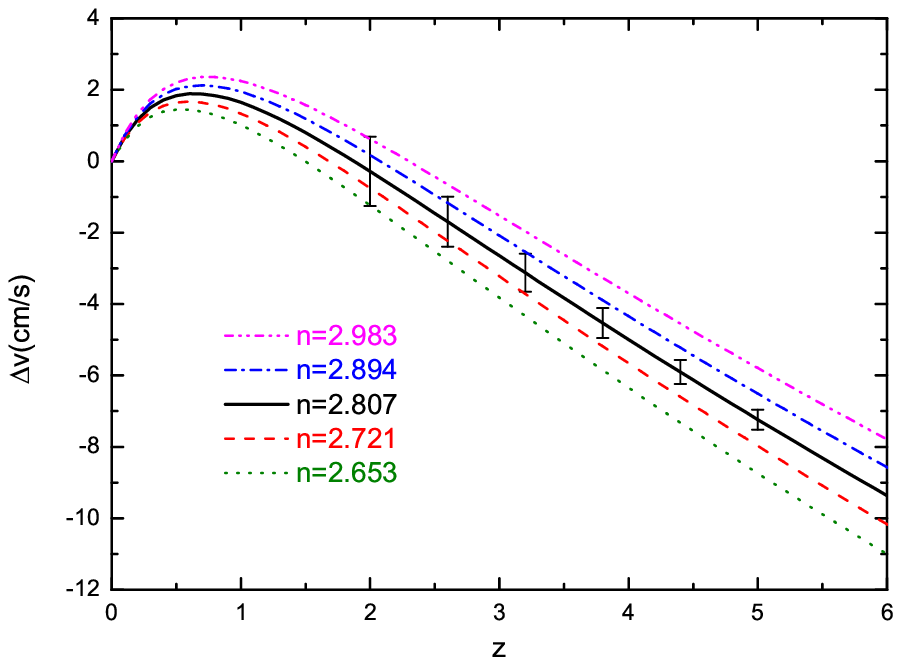} \\
\end{array}$
\caption{The SL test for the NADE model. In the left panel, the
$\Lambda$CDM model with $\Omega_{\rm m0}=0.274$ is used to be the
fiducial model; in the right panel, the NADE model with $n=2.807$ is
taken as the fiducial model.}\label{fig:NADE}
\end{figure}



\begin{figure}[htbp]
\centering $\begin{array}{c@{\hspace{0.2in}}c}
\multicolumn{1}{l}{\mbox{}} &
\multicolumn{1}{l}{\mbox{}} \\
\includegraphics[scale=0.8]{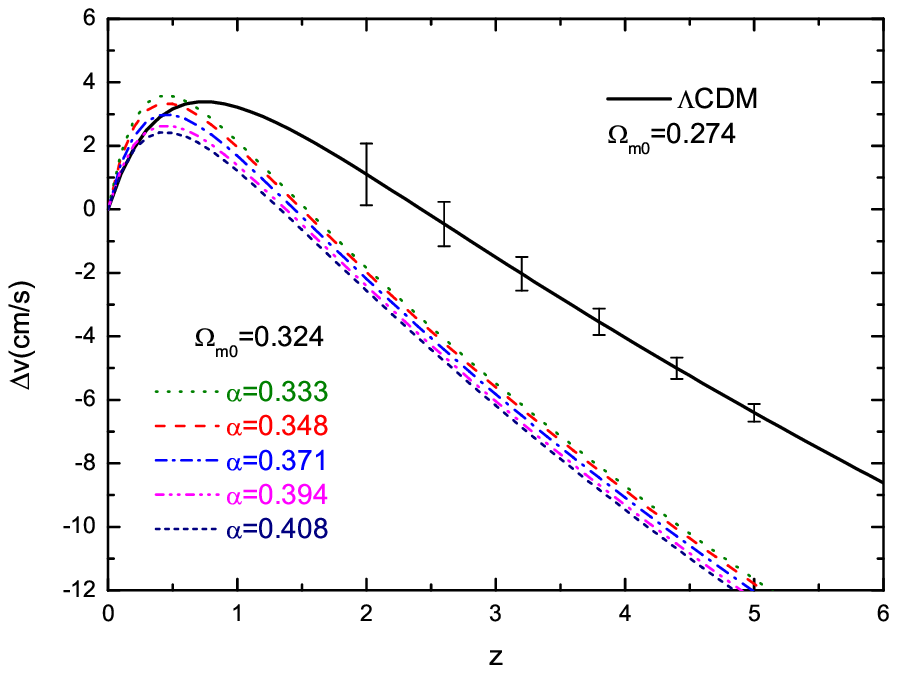} &\includegraphics[scale=0.8]{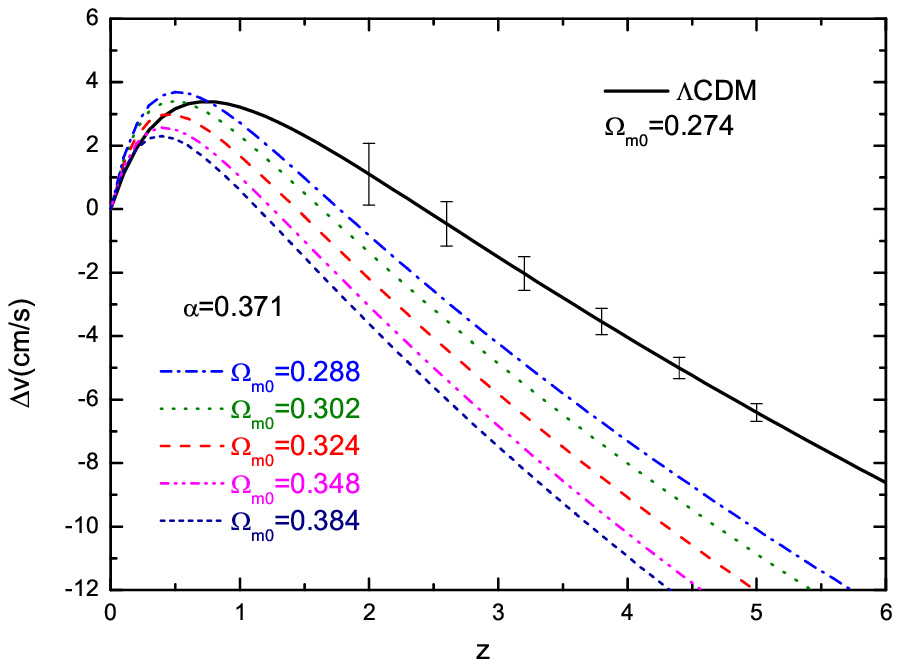} \\
\end{array}$
\caption{The SL test for the RDE model. The $\Lambda$CDM model with
$\Omega_{\rm m0}=0.274$ is used to be the fiducial model. In the
left panel, we fix $\Omega_{\rm m0}=0.324$ and vary $\alpha$; in the
right panel, we fix $\alpha=0.371$ and vary $\Omega_{\rm m0}$.}
\label{fig:RDE1}
\end{figure}



\begin{figure}[htbp]
\centering $\begin{array}{c@{\hspace{0.2in}}c}
\multicolumn{1}{l}{\mbox{}} &
\multicolumn{1}{l}{\mbox{}} \\
\includegraphics[scale=0.8]{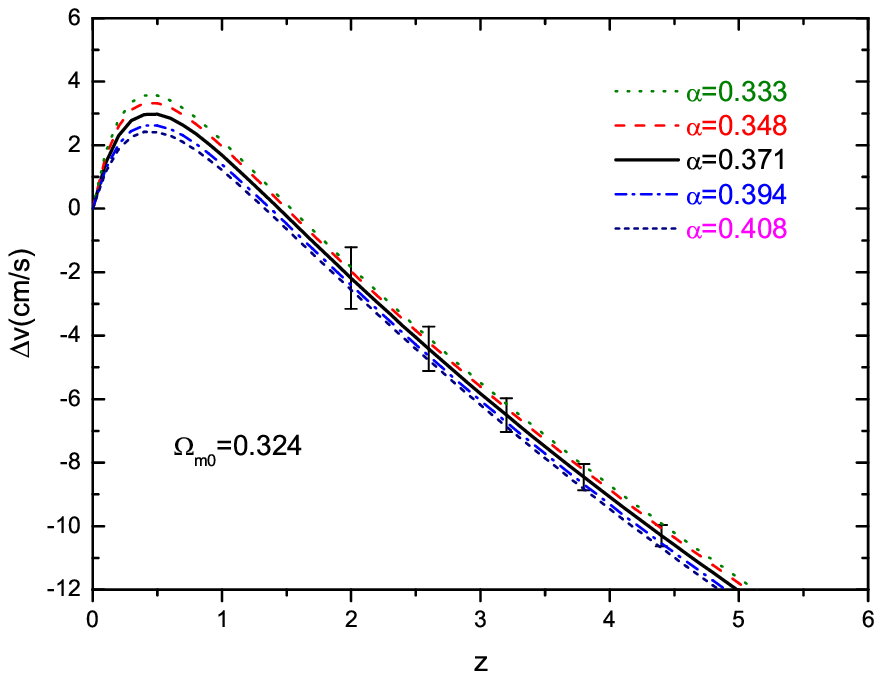} &\includegraphics[scale=0.8]{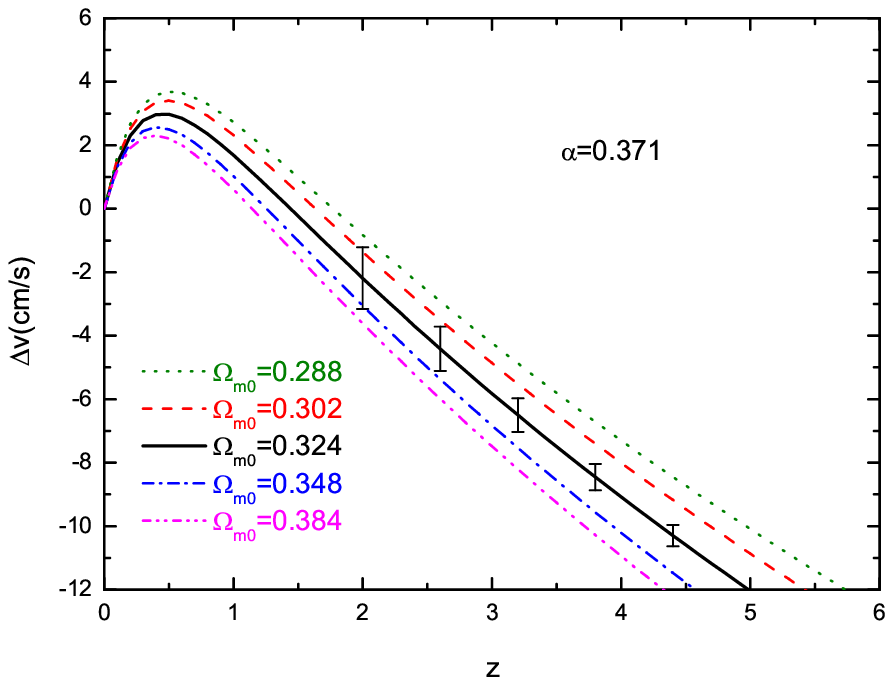} \\
\end{array}$
\caption{The SL test for the RDE model. The RDE model with
$\Omega_{\rm m0}=0.324$ and $\alpha=0.371$ is used to be the
fiducial model. In the left panel, we fix $\Omega_{\rm m0}=0.324$
and vary $\alpha$; in the right panel, we fix $\alpha=0.371$ and
vary $\Omega_{\rm m0}$.} \label{fig:RDE2}
\end{figure}


First, we discuss the velocity shift behavior in the new agegraphic
dark energy (NADE) model. Numerically solving the differential
equation (\ref{keyq}) and inserting Eq.~(\ref{Ez}) into
Eq.~(\ref{deltav}), one can reconstruct the $\Delta v(z)$ curves for
the NADE model. Note that the NADE model is a single-parameter
model, and the sole model parameter is $n$. The current cosmological
constraint on the parameter $n$ is $n=2.807^{+0.087}_{-0.086}$
(1$\sigma$) $^{+0.176}_{-0.170}$ (2$\sigma$) \cite{holos09}. In this
work, we take the values of $n$ as the central value as well as the
one-sigma and two-sigma limits of the above observational
constraint. In Fig.~\ref{fig:NADE}, we reconstruct the velocity
shift behavior of the NADE model. In the left panel, we use the
$\Lambda$CDM model as a fiducial model to perform an SL test. For
the $\Lambda$CDM cosmology, $\Omega_{\rm m0}$ is chosen to be 0.274
as given by the WMAP five-year observations \cite{Komatsu:2008hk}.
Note also that the Hubble constant $H_0$ in Eq.~(\ref{deltav}) is
taken as 72 km s$^{-1}$ Mpc$^{-1}$, in the whole discussion. From
the left panel of Fig.~\ref{fig:NADE} we see that the NADE model can
be distinguished from the $\Lambda$CDM model via the SL test.
However, it is interesting to notice that the curve with $n=0.2894$
still lies in the 1$\sigma$ error range of the $\Lambda$CDM fiducial
model within the redshift desert $2\lesssim z\lesssim 5$. So, it is
fair to say that actually the SL test could not completely
distinguish the NADE model from the $\Lambda$CDM model, though it
can do it rather effectively. Of course, if the SL test is combined
with the low-redshift observations such as the SNIa, weak lensing,
and BAO, the NADE model can be completely distinguished from the
$\Lambda$CDM model. On the other hand, we also perform an SL test by
using the NADE model with $n=2.807$ (the central value of $n$ given
by the current observational constraint) as a fiducial model. We
show this case in the right panel of Fig.~\ref{fig:NADE}. In this
case, we see that the prospective SL test is very powerful to be
used to constrain the NADE model, and it is clearly better than the
current low-redshift observations.

Next, we switch to the Ricci dark energy (RDE) model. For the RDE
model, the Hubble expansion history is described by Eq.~(\ref{Ea}),
so the $\Delta v(z)$ behavior can be reconstructed by substituting
Eq.~(\ref{Ea}) into Eq.~(\ref{deltav}). The RDE model has two model
parameters, $\alpha$ and $\Omega_{\rm m0}$, and the current
observational constraint results are \cite{holos09}:
$\alpha=0.371^{+0.023}_{-0.023}$ ($1\sigma$) $^{+0.037}_{-0.038}$
($2\sigma$) and $\Omega_{\rm m0}=0.324^{+0.024}_{-0.022}$
($1\sigma$) $^{+0.040}_{-0.036}$ ($2\sigma$). Like the above
discussion, we will still first employ the standard dark energy
cosmology, the $\Lambda$CDM model, as a fiducial model to perform an
SL test for the RDE model. For the fiducial $\Lambda$CDM model, we
still take $\Omega_{\rm m0}=0.274$, the value given by WMAP
\cite{Komatsu:2008hk}. In this test, one can forecast how well a
deviation of the RDE model from the $\Lambda$CDM model can be
detected. We show the results of such an SL test in
Fig.~\ref{fig:RDE1}. In the left panel, the parameter $\Omega_{\rm
m0}$ is fixed to be 0.324 for the RDE model, and the parameter
$\alpha$ of the RDE model is adjustable; in the right panel, we fix
$\alpha=0.371$ and vary $\Omega_{\rm m0}$ for the RDE model. From
this figure, it is clear to see that the SL test in the ``redshift
desert'' is very successful in distinguishing the RDE model from the
standard $\Lambda$CDM cosmology. The two models can be distinguished
via the SL test at about $6-8$ $\sigma$ level, assuming only a
10-year survey. This result is reasonable and understandable, since
the RDE model has a tracker behavior \cite{Zhang:2009un} so that its
matter-dominated era is different from that of the $\Lambda$CDM
model, and the SL test is just the best way to probe the behavior of
the matter-dominated phase of dark energy models. Furthermore, it is
worth noticing that such measurements are mostly sensitive to the
matter density $\Omega_{\rm m0}$, while the dependence on $\alpha$
is weaker. To forecast whether the SL test is able to break the
degeneracy of the parameters of the RDE model, we further use the
RDE model with $\Omega_{\rm m0}=0.324$ and $\alpha=0.371$ (the
central values of the current observational constraint) as the
fiducial model to perform an SL test. The results of this test are
plotted in Fig.~\ref{fig:RDE2}. In the left panel, we fix
$\Omega_{\rm m0}=0.324$ and vary $\alpha$ in the 2 $\sigma$ range of
the current observational result; in the right panel, we fix
$\alpha=0.371$ and vary $\Omega_{\rm m0}$ in the 2 $\sigma$ range of
the current observational result. From Fig.~\ref{fig:RDE2}, we can
see that the SL test with a 10-year survey could not provide
precision determination of the parameter $\alpha$, but could
precisely determine the value of $\Omega_{\rm m0}$ for the RDE
model. This result is in agreement with the previous work on the SL
test \cite{Corasaniti:2007bg}; in Ref.~\cite{Corasaniti:2007bg} it
is shown that the Sandage-Loeb constraints on $w$ are not
competitive with those inferred from other low-redshift
observations. The equation of state $w$ of RDE is mainly determined
by the parameter $\alpha$, so the SL test could not constrain the
value of $\alpha$ precisely. However, as pointed out in
Ref.~\cite{Corasaniti:2007bg}, one should note that the constraints
obtained by SL test decrease linearly with time, so for measurements
made over a century, and with the expected larger number of QSOs,
the SL limits on $w$ can easily be at the few percent level.

\begin{figure}[htbp]
\begin{center}
\includegraphics[scale=0.8]{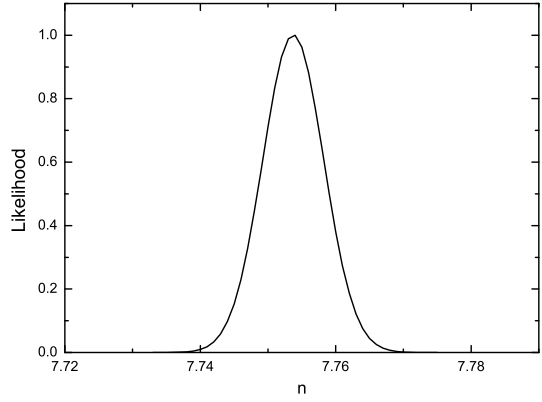}
\caption{Predicted likelihood distribution of the parameter $n$ of
the NADE model from the SL test.}\label{fig:adelikelihood}
\end{center}
\end{figure}

\begin{figure}[htbp]
\begin{center}
\includegraphics[scale=0.7]{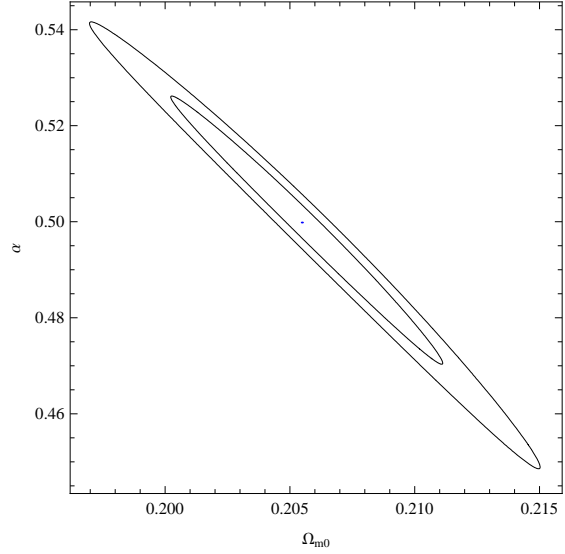}
\caption{Predicted probability contours at $68.3\%$ and $95.4\%$
confidence levels in the ($\Omega_{m0}$, $\alpha$) plane for the RDE
model from the SL test.}\label{fig:rdecontour}
\end{center}
\end{figure}

Finally, let us show how well the proposed SL test could constrain
the two dark energy models by using a Fisher matrix method. To find
the expected precision of the SL test with CODEX, one must assume a
fiducial model, and then simulate the experiment assuming it as a
reference model. We employ the $\Lambda$CDM model as the fiducial
model and produce the mock data of the velocity-drift in the
redshift desert with the error bars given by Eq.~(\ref{error}).
Next, we will perform a Fisher matrix analysis on the model
parameter space with the above assumption. The $\chi^2$ of the
analysis is given by
\begin{equation}
\chi^2_{SL}=\sum\limits_{i=1}^{240}{[\Delta v^{\rm
model}(z_i)-\Delta v^{\rm data}(z_i)]^2 \over\sigma_{\Delta
v}^2(z_i)},
\end{equation}
where $\Delta v^{\rm data}(z_i)$ denotes the mock data produced by
the fiducial model, $\Delta v^{\rm model}(z_i)$ is the theoretical
prediction of the dark energy model under investigation, and
$\sigma_{\Delta v}^2(z_i)$ is the error bar estimated from
Eq.~(\ref{error}). The NADE model is a single-parameter model, so
once the parameter $n$ is given, the parameter $\Omega_{\rm m0}$ as
well as other properties and dynamical evolution will be determined
accordingly. In Fig.~\ref{fig:adelikelihood} we plot the likelihood
distribution of the parameter $n$ of the NADE model as expected from
the SL test. We see that the parameter $n$ is able to be accurately
determined by the SL test, $n=7.754^{+0.004}_{-0.005}$ (1$\sigma$).
So, the SL test could obtain $\sigma_n\simeq 0.005$, better than the
current observational result by at least one order of magnitude. The
RDE model has two model parameters, $\Omega_{\rm m0}$ and $\alpha$.
In Fig.~\ref{fig:rdecontour} we plot the 1$\sigma$ and 2$\sigma$
contours in the $\Omega_{\rm m0}-\alpha$ plane for the RDE model.
The 1$\sigma$ results are: $\Omega_{\rm
m0}=0.2055^{+0.0056}_{-0.0052}$ and
$\alpha=0.5000^{+0.0262}_{-0.0297}$. So, the SL test would obtain
$\sigma_{\Omega_{\rm m0}}\simeq 0.005$, better than the current
constraint by one order of magnitude, and $\sigma_{\alpha}\simeq
0.03$, at the same accuracy level compared to the current
observational result. The above analysis reinforces the conclusion
that the two dark energy models indeed can be constrained by the SL
measurements with high significance.

\section{Concluding remarks}\label{sec:conclusion}

The Sandage-Loeb test is a promising method for constraining dark
energy by using the direct measurements of the temporal shift of the
quasar Lyman-$\alpha$ absorption lines at high redshift ($2\lesssim
z\lesssim 5$). While the signal of this effect is extremely small,
the near-future large telescopes with ultrahigh resolution
spectrographs (such as the CODEX under design) will definitely be
capable of measuring such a small signal over a period as short as
ten years. Notwithstanding, one still may ask whether it is
worthwhile to probe $z\gtrsim 2$, since in the standard dark energy
cosmology such as the $\Lambda$CDM model or the slowly rolling
scalar field model the dark energy is subdominant at redshift
$z\gtrsim 1$ and almost completely negligible at $z\gtrsim 2$. The
answer to this question is affirmative. As explained in
Ref.~\cite{Corasaniti:2007bg}, it is quite rational to look for the
signatures of dark energy at all available epochs, because at
present we do not know much about the physical nature and
cosmological origin of dark energy, and there are too many
possibilities for dark energy. Especially, there are lots of dark
energy models in which dark energy density is non-negligible at high
redshift. So, it is meaningful to study the future observations of
velocity shift and their impact on dark energy models. The SL test
on Chaplygin gas model and interacting dark energy model has been
studied in Ref.~\cite{Corasaniti:2007bg}. In
Ref.~\cite{Balbi:2007fx}, more classes of dark energy models have
been investigated with the SL test. In particular, the original
holographic dark energy model has been explored with the SL test in
Ref.~\cite{Zhang:2007zg}. So, the exploration of the other two
holographic dark energy models, the new agegraphic dark energy model
and the Ricci dark energy model, is naturally the next step, and
this will provide a complementary to the work of
Ref.~\cite{Zhang:2007zg} and keep the investigation of holographic
scenarios of dark energy more complete.

In this paper, we have analyzed the prospects for constraining the
new agegraphic dark energy model and the Ricci dark energy model at
the redshift desert $2\lesssim z\lesssim 5$ from the Sandage-Loeb
test. The NADE model is a sole-parameter model; the parameter $n$
together with an initial condition can determine the whole
cosmological evolution history of the NADE model. Actually, the
evolution of this model is similar to that of a slowly rolling
scalar field. Thus, though the SL test can be used to distinguish
the NADE model from the $\Lambda$CDM model, this discrimination is
not absolute. It is obvious that if the SL test is combined with the
low-redshift observations such as the SNIa, weak lensing, and BAO,
the NADE model can be completely distinguished from the $\Lambda$CDM
model. Furthermore, we forecast that the prospective SL test is very
powerful to constrain the NADE model, and it is better than the
current low-redshift observations. The RDE model is a two-parameter
model, and it has a tracking solution so that it is very different
from the $\Lambda$CDM model in the matter-dominated epoch. So, the
SL test is very successful in distinguishing the RDE model from the
$\Lambda$CDM model. The two models can be distinguished via the SL
test at about $6-8$ $\sigma$ level, assuming only a 10-year survey.
As the SL test mostly probes the matter density at high redshift,
the constraint on $\Omega_{\rm m0}$ is very strong, while the
constraint on $\alpha$ is much weaker.

\begin{acknowledgments}
We are grateful to the referee for the insightful comments and
suggestions, which have led to significant improvement to this
paper. We also thank Dr. Hongbao Zhang for useful discussions. This
work was supported by the Natural Science Foundation of China under
Grant Nos.~10705041 and 10975032.
\end{acknowledgments}


\begin{thebibliography}{}

\bibitem{Riess98}
A.~G.~Riess {\it et al.} [Supernova Search Team Collaboration],
  Astron.\ J.\  {\bf 116}, 1009 (1998)
  [arXiv:astro-ph/9805201];
S.~Perlmutter {\it et al.} [Supernova Cosmology Project
Collaboration], Astrophys.\ J.\  {\bf 517}, 565 (1999)
  [arXiv:astro-ph/9812133].

\bibitem{Tegmark04}
M.~Tegmark {\it et al.} [SDSS Collaboration],
    Phys.\ Rev.\  D {\bf 69}, 103501 (2004)
  [arXiv:astro-ph/0310723];
K.~Abazajian {\it et al.} [SDSS Collaboration],
  Astron.\ J.\  {\bf 128}, 502 (2004)
  [arXiv:astro-ph/0403325].

\bibitem{Spergel03}
D.~N.~Spergel {\it et al.} [WMAP Collaboration],
  Astrophys.\ J.\ Suppl.\  {\bf 148}, 175 (2003)
  [arXiv:astro-ph/0302209];
C.~L.~Bennett {\it et al.} [WMAP Collaboration],
  Astrophys.\ J.\ Suppl.\  {\bf 148}, 1 (2003)
  [arXiv:astro-ph/0302207].


\bibitem{dark energy}
V.~Sahni and A.~A.~Starobinsky,
  Int.\ J.\ Mod.\ Phys.\  D {\bf 9} (2000) 373
  [arXiv:astro-ph/9904398];
S.~M.~Carroll,
  Living Rev.\ Rel.\  {\bf 4} (2001) 1
  [arXiv:astro-ph/0004075];
P.~J.~E.~Peebles and B.~Ratra,
  Rev.\ Mod.\ Phys.\  {\bf 75} (2003) 559
  [arXiv:astro-ph/0207347];
T.~Padmanabhan,
  Phys.\ Rept.\  {\bf 380} (2003) 235
  [arXiv:hep-th/0212290];
E.~J.~Copeland, M.~Sami and S.~Tsujikawa,
  Int.\ J.\ Mod.\ Phys.\  D {\bf 15} (2006) 1753
  [arXiv:hep-th/0603057];
  E.~V.~Linder,
  Am.\ J.\ Phys.\  {\bf 76}, 197 (2008)
  [arXiv:0705.4102 [astro-ph]].



\bibitem{Sandage}
  A.\ Sandage,  Astrophys. J. {\bf 139}, 319 (1962).

\bibitem{Loeb}
  A.\ Loeb,  Astrophys. J. {\bf 499}, L111 (1998) [arXiv:astro-ph/9802122].

\bibitem{Corasaniti:2007bg}
  P.~S.~Corasaniti, D.~Huterer and A.~Melchiorri,
  Phys.\ Rev.\  D {\bf 75}, 062001 (2007)
  [arXiv:astro-ph/0701433].


\bibitem{Balbi:2007fx}
  A.~Balbi and C.~Quercellini,
  Mon. Not. Roy. Astro. Soc. {\bf 382}, 1623 (2007)
  [arXiv:0704.2350 [astro-ph]].



\bibitem{Zhang:2007zg}
  H.~Zhang, W.~Zhong, Z.~H.~Zhu and S.~He,
  Phys.\ Rev.\  D {\bf 76}, 123508 (2007)
  [arXiv:0705.4409 [astro-ph]].

\bibitem{Li:2004rb}
  M.~Li,
  Phys.\ Lett.\ B {\bf 603}, 1 (2004)
  [arXiv:hep-th/0403127].

\bibitem{nade}
  H.~Wei and R.~G.~Cai,
  Phys.\ Lett.\  B {\bf 660}, 113 (2008)
  [arXiv:0708.0884 [astro-ph]].

\bibitem{rde}
  C.~Gao, F.~Q. Wu, X.~Chen and Y.~G.~Shen,
  Phys.\ Rev.\ D {\bf 79}, 043511 (2009)
  [arXiv:0712.1394 [astro-ph]].


\bibitem{holop}
  G.~'t Hooft,
  arXiv:gr-qc/9310026;
L.~Susskind,
  J.\ Math.\ Phys.\  {\bf 36}, 6377 (1995)
  [arXiv:hep-th/9409089].

\bibitem{Cohen99}
A.~G.~Cohen, D.~B.~Kaplan and A.~E.~Nelson,
  Phys.\ Rev.\ Lett.\  {\bf 82}, 4971 (1999)
  [arXiv:hep-th/9803132].

\bibitem{Hsu04}
S.~D.~H.~Hsu,
  Phys.\ Lett.\  B {\bf 594}, 13 (2004)
  [arXiv:hep-th/0403052].

\bibitem{holoext}
  Q.~G.~Huang and M.~Li,
  JCAP {\bf 0408}, 013 (2004)
  [arXiv:astro-ph/0404229];
  Q.~G.~Huang and M.~Li,
  JCAP {\bf 0503}, 001 (2005)
  [arXiv:hep-th/0410095];
  X.~Zhang,
  Int.\ J.\ Mod.\ Phys.\  D {\bf 14}, 1597 (2005)
  [arXiv:astro-ph/0504586];
  X.~Zhang,
  Phys.\ Lett.\  B {\bf 648}, 1 (2007)
  [arXiv:astro-ph/0604484];
  X.~Zhang,
  Phys.\ Rev.\  D {\bf 74}, 103505 (2006)
  [arXiv:astro-ph/0609699];
  J.~Zhang, X.~Zhang and H.~Liu,
  Phys.\ Lett.\  B {\bf 651}, 84 (2007)
  [arXiv:0706.1185 [astro-ph]];
  Y.~Z.~Ma and X.~Zhang,
  Phys.\ Lett.\  B {\bf 661}, 239 (2008)
  [arXiv:0709.1517 [astro-ph]];
  X.~Wu and Z.~H.~Zhu,
  Phys.\ Lett.\  B {\bf 660}, 293 (2008)
  [arXiv:0712.3603 [astro-ph]];
  X.~Zhang,
  Phys.\ Lett.\  B {\bf 683}, 81 (2010)
  [arXiv:0909.4940 [gr-qc]];
  M.~R.~Setare, J.~Zhang and X.~Zhang,
  JCAP {\bf 0703}, 007 (2007)
  [arXiv:gr-qc/0611084];
  J.~Zhang, X.~Zhang and H.~Liu,
  Eur.\ Phys.\ J.\  C {\bf 52}, 693 (2007)
  [arXiv:0708.3121 [hep-th]];
  J.~Zhang, X.~Zhang and H.~Liu,
  Phys.\ Lett.\  B {\bf 659}, 26 (2008)
  [arXiv:0705.4145 [astro-ph]];
  Y.~G.~Gong,
  Phys.\ Rev.\  D {\bf 70}, 064029 (2004)
  [arXiv:hep-th/0404030];
  B.~Wang, E.~Abdalla and R.~K.~Su,
  Phys.\ Lett.\  B {\bf 611}, 21 (2005)
  [arXiv:hep-th/0404057];
  X.~Wu, R.~G.~Cai and Z.~H.~Zhu,
  Phys.\ Rev.\  D {\bf 77}, 043502 (2008)
  [arXiv:0712.3604 [astro-ph]];
  M.~Li, C.~Lin and Y.~Wang,
  JCAP {\bf 0805}, 023 (2008)
  [arXiv:0801.1407 [astro-ph]];
  M.~Li, X.~D.~Li, C.~Lin and Y.~Wang,
  Commun. Theor. Phys. {\bf 51}, 181 (2009)
  [arXiv:0811.3332 [hep-th]];
  B.~Wang, Y.~G.~Gong and E.~Abdalla,
  Phys.\ Lett.\  B {\bf 624}, 141 (2005)
  [arXiv:hep-th/0506069];
  B.~Wang, C.~Y.~Lin and E.~Abdalla,
  Phys.\ Lett.\  B {\bf 637}, 357 (2006)
  [arXiv:hep-th/0509107];
  B.~Wang, C.~Y.~Lin, D.~Pavon and E.~Abdalla,
  Phys.\ Lett.\  B {\bf 662}, 1 (2008)
  [arXiv:0711.2214 [hep-th]];
  S.~Nojiri and S.~D.~Odintsov,
  Gen.\ Rel.\ Grav.\  {\bf 38}, 1285 (2006)
  [arXiv:hep-th/0506212].

\bibitem{holos09}
  M.~Li, X.~D.~Li, S.~Wang and X.~Zhang,
  JCAP {\bf 0906}, 036 (2009)
  [arXiv:0904.0928 [astro-ph.CO]].



\bibitem{holofitzx}
  X.~Zhang and F.~Q.~Wu,
  Phys.\ Rev.\  D {\bf 72}, 043524 (2005)
  [arXiv:astro-ph/0506310];
  X.~Zhang and F.~Q.~Wu,
  Phys.\ Rev.\  D {\bf 76}, 023502 (2007)
  [arXiv:astro-ph/0701405];
  Q.~G.~Huang and Y.~G.~Gong,
  JCAP {\bf 0408}, 006 (2004)
  [arXiv:astro-ph/0403590];
  Z.~Chang, F.~Q.~Wu and X.~Zhang,
  Phys.\ Lett.\  B {\bf 633}, 14 (2006)
  [arXiv:astro-ph/0509531];
  J.~Y.~Shen, B.~Wang, E.~Abdalla and R.~K.~Su,
  Phys.\ Lett.\  B {\bf 609}, 200 (2005)
  [arXiv:hep-th/0412227];
  Z.~L.~Yi and T.~J.~Zhang,
  Mod.\ Phys.\ Lett.\  A {\bf 22}, 41 (2007)
  [arXiv:astro-ph/0605596];
  Y.~Z.~Ma, Y.~Gong and X. Chen,
  Eur.\ Phys.\ J.\  C {\bf 60}, 303 (2009)
  [arXiv:0711.1641 [astro-ph]];
  Q.~Wu, Y.~Gong, A.~Wang and J.~S.~Alcaniz,
  Phys.\ Lett.\  B {\bf 659}, 34 (2008)
  [arXiv:0705.1006 [astro-ph]];
  M.~Li, X.~D.~Li, S.~Wang, Y.~Wang and X.~Zhang,
  JCAP {\bf 0912}, 014 (2009)
  [arXiv:0910.3855 [astro-ph.CO]];
  M.~Li, X.~D.~Li and X.~Zhang,
  arXiv:0912.3988 [astro-ph.CO].

\bibitem{nadeext}
H.~Wei and R.~G.~Cai,
 Phys.\ Lett.\  B {\bf 663}, 1 (2008) [arXiv:0708.1894];
  I.~P.~Neupane,
  Phys.\ Rev.\  D {\bf 76}, 123006 (2007)
  [arXiv:0709.3096 [hep-th]];
  J.~Zhang, X.~Zhang and H.~Liu,
  Eur.\ Phys.\ J.\  C {\bf 54}, 303 (2008)
  [arXiv:0801.2809 [astro-ph]];
  Y.~W.~Kim, H.~W.~Lee, Y.~S.~Myung and M.~I.~Park,
  Mod.\ Phys.\ Lett.\  A {\bf 23}, 3049 (2008)
  [arXiv:0803.0574 [gr-qc]];
  J.~P.~Wu, D.~Z.~Ma and Y.~Ling,
  Phys.\ Lett.\  B {\bf 663}, 152 (2008)
  [arXiv:0805.0546 [hep-th]];
  J.~Cui, L.~Zhang, J.~Zhang and X.~Zhang,
  Chin.\ Phys.\  B {\bf 19}, 019802 (2010)
  [arXiv:0902.0716 [astro-ph.CO]].
  L.~Zhang, J.~Cui, J.~Zhang and X.~Zhang,
  Int.\ J.\ Mod.\ Phys.\  D {\bf 19}, 21 (2010)
  [arXiv:0911.2838 [astro-ph.CO]];
  X.~L.~Liu and X.~Zhang,
  Commun.\ Theor.\ Phys.\  {\bf 52}, 761 (2009)
  [arXiv:0909.4911 [astro-ph.CO]];
  X.~L.~Liu, J.~Zhang and X.~Zhang,
  Phys.\ Lett.\  B {\bf 689}, 139 (2010)
  [arXiv:1005.2466 [gr-qc]].




\bibitem{Zhang:2009un}
  X.~Zhang,
  Phys.\ Rev.\  D {\bf 79}, 103509 (2009)
  [arXiv:0901.2262 [astro-ph.CO]].




\bibitem{ricciext}
  C.~J.~Feng and X.~Zhang,
  Phys.\ Lett.\  B {\bf 680}, 399 (2009)
  [arXiv:0904.0045 [gr-qc]];
  C.~J.~Feng,
  arXiv:0806.0673 [hep-th];
  C.~J.~Feng,
  Phys.\ Lett.\  B {\bf 670}, 231 (2008)
  [arXiv:0809.2502 [hep-th]];
  L.~Xu, W.~Li and J.~Lu,
  Mod.\ Phys.\ Lett.\  A {\bf 24}, 1355 (2009)
  [arXiv:0810.4730 [astro-ph]];
  C.~J.~Feng,
  Phys.\ Lett.\  B {\bf 676}, 168 (2009)
  [arXiv:0812.2067 [hep-th]];
  K.~Y.~Kim, H.~W.~Lee and Y.~S.~Myung,
  arXiv:0812.4098 [gr-qc].

\bibitem{Cai:2007us}
  R.~G.~Cai,
  Phys.\ Lett.\  B {\bf 657}, 228 (2007)
  [arXiv:0707.4049 [hep-th]].

\bibitem{Cai:2008nk}
  R.~G.~Cai, B.~Hu and Y.~Zhang,
  Commun. Theor. Phys. {\bf 51}, 954 (2009)
  [arXiv:0812.4504 [hep-th]].

\bibitem{Pasquini05}
L.~Pasquini {\it et al.}, Proc. Int. Astron. Union {\bf 1}, 193
(2005).

\bibitem{Komatsu:2008hk}
  E.~Komatsu {\it et al.}  [WMAP Collaboration],
  Astrophys.\ J.\ Suppl.\  {\bf 180}, 330 (2009)
  [arXiv:0803.0547 [astro-ph]].































\end{thebibliography}
\end{document}